# The gradual transformation of inland areas - human plowing, horse plowing and equity incentives

Hongfa Zi[1]

Qinghai Nationalities University, School of Economics and Trade, Qinghai Province, China

Many modern areas have not learned their lessons and often hope for the wisdom of later generations, resulting in them only possessing modern technology and difficult to iterate ancient civilizations. At present, there is no way to tell how we should learn from history and promote the gradual upgrading of civilization. Therefore, we must tell the history of civilization's progress and the means of governance, learn from experience to improve the comprehensive strength and survival ability of civilization, and achieve an optimal solution for the tempering brought by conflicts and the reduction of internal conflicts. Firstly, we must follow the footsteps of history and explore the reasons for the long-term stability of each country in conflict, including providing economic benefits to the people and means of suppressing them; then, use mathematical methods to demonstrate how we can achieve the optimal solution at the current stage. After analysis, we can conclude that the civilization transformed from human plowing to horse plowing can easily suppress the resistance of the people and provide them with the ability to resist; The selection of rulers should consider multiple institutional aspects, such as technology force exams, elections, and drawing lots; Economic development follows a lognormal distribution and can be adjusted by standard deviation and the number of virtual employees. Using a lognormal distribution with the maximum value to divide shareholding can adjust the wealth gap.

* Corresponding author: 1697358179@qq.com

## Introduction

The development of human civilization has always been accompanied by conflicts and changes, but not all societies can learn from history. Many civilizations, despite possessing modern technology, have stagnated due to outdated governance models and are essentially "ancient societies cloaked in technology" (Vaesen, Krist, et al). A long-standing social problem that has never been fixed comes from interest conflicts between the upper and lower classes whose reasons are known. Since lacking real solutions, managers resort to deception to sustain the organization, yet grey rhino risks remain unaddressed and demand settlement. To promote the civilization upgrade and resolve conflicts, it is necessary to systematically study historical laws, optimize governance strategies using mathematical tools, and ultimately achieve a dual improvement in comprehensive strength and stability. Specifically, this process can be divided into three key stages: historical experience summary, governance model optimization, and dynamic balance between conflict and development.

Long term stable civilizations often rely on two core means: the distribution of economic benefits and the maintenance of power suppression. Taking ancient China as an example, Confucianism incorporated the elite class into the ruling system through the imperial examination system, but officials with imperial examination backgrounds were still corrupt. The excessive tax interception and corvee caused by corruption far exceeded normal taxes, and peasant uprisings led to the collapse of the country (Chan, Kenneth, and Jean). Creating morality and stability for tax revenue cannot deliver genuine stability; evading responsibility means forsaking long-term stability. The collapse of the Roman Empire was partly due to financial collapse and military loss of control, resulting in the dual failure of economic incentives and violent monopolies. The progress of production modes also affects social stability - the civilization of horse plowing is more efficient than manual farming, as it can increase economic output and strengthen suppression capabilities by controlling cavalry. Therefore, the first step in upgrading civilization is to identify the critical point of economic and power balance in history.

The design of civilized systems needs to be adapted to civilized systems through quantitative analysis. In the selection of power holders, a single mechanism (such as pure elections or hereditary) can easily lead to oligarchic rule and class solidification, while a mixed system (elections+lots) may be better (Reynolds, Andrew, and Marco). Exams are essentially just a type of ranking, so we should combine elections, force exams, and lotteries. For example, the Republic of Venice avoids family monopolies through the "Grand council drawing system", while the imperial examination system selects capable individuals through examinations, and the combination of the two can be optimized through models. At the economic level. Lognormal distribution can simulate wealth distribution: the majority of people are concentrated in the middle-income group, with a few elites occupying high positions. By setting a maximum threshold and standard deviation for equity distribution, the destructive nature of wealth inequality can be suppressed.

Conflict is both a threat to civilization and a catalyst for progress. The pressure of war may force technological breakthroughs (such as the industrial revolution spurred by modern military competition in Europe), but excessive internal class contradictions can lead to collapse. The optimal solution lies in finding the "conflict threshold" - through history, it can be found that certain civilizations have the highest reform efficiency during rebellions (such as the Glorious

Revolution in England). At the same time, the upgrading of civilization needs to be promoted in stages: from manpower to horse plowing, and then to mechanization, each step needs to be matched with institutional adjustments. For example, during the horse plowing era, it was necessary to strengthen the central cavalry to suppress local areas, while during the industrial era, it was necessary to allocate resources reasonably to avoid revolution. The ultimate goal is to make conflicts a "controllable quenching" through dynamic adjustment, rather than a systemic disaster.

In summary, history is not a cycle, but a database that can be parsed; Governance is not art, but an optimizable algorithm. By extracting historical parameters (economic power balance point), constructing mathematical models (selection system, wealth distribution), and dynamically regulating conflict levels, civilization can gradually approach the "optimal state" under current technological conditions. This framework is not only applicable to countries, but also to enterprises or international organizations - after all, the essence of all collective existence is the precise calculation of interest distribution and power dynamics. Social science is a young science, but it can only use the humanity of other people to restrain humanity and not transform it.

## Literature review

### Short sighted and corrupt people farming civilization

The rule in human civilization is far more important than production, so its agriculture is even more backward. Potatoes, sweet potatoes, cassava, corn, yacon, and maca are all products of the American continent. Did the Eurasian continent, as the mainstream civilization of ancient times, not have tuber crops? No, even in China, there are big-yam-bean (air potatoes can be harvested continuously for over a decade without the need for a long period of germination), yam (Chinese yam), kudzu root, gastrodia elata, big radix asparagi, and polygonum multiflorum; India also has crops such as amorphophallus paeoniifolius taro that contain starch (Sanderson). Some sweet potatoes planting and storage techniques were artificially lost because they were not favored by landlords led by the emperor. Bureaucrats and examination-system landlords preferred to annex the land and other means of production of the people during times of famine and let them engage in productive labor (Perkins, Dwight). However, this method only creates potentially violent poor people, examinees who merely learn knowledge, and the vulnerable oppressed by external enemies, without yielding many benefits. Examinees who only learn knowledge without creating it are not as excellent and easy as scholars who create problems and create truth knowledge. If taking the mathematical limit as imperial examination approaches infinity, are examiners qualified to set questions for the emperor? Because it does not conform to natural forms, so votes and force exams are more popular among the supervision. Bureaucrats who cannot supervise each other cannot understand this truth, and corrupt and short-sighted officers do not even know how much proportion of military pay lower level officers will give to their subordinates (Swope, Kenneth). Due to the lack of cavalry to crush tax-free landlords with numerous families, some corrupt bureaucrats had to continue to increase their taxation efforts, instead prompting the people to sell their land to bureaucrats and landlords to avoid paying grain taxes (Deng, Kent Gang). The bureaucratic group can be regarded as an enterprise with no shareholders holding its equity (equity dividends and add-value of public property still exist), so the only way to mutually supervise officials is through common interests (only officials can deal with officials), so a certain percentage of taxes should be converted into salary and equity dividends and distributed proportionally. For example, one half or one third of the tax revenue equals the all payroll.

**The horse plowing civilization that brings resistance and civilization evolution**

The farming efficiency of horses and the killing efficiency of cavalry far exceed that of humans, so they can further transform. Both ancient Europe and ancient China had horse plowing civilizations, but their attitudes towards the two were completely different. The landlord class in ancient China severely exploited others, and could not tolerate constant roaming by outstanding cavalry who resisted. There were civil service examinations, but no corresponding force or cavalry examinations. Until the Song and Ming dynasties, the breeding and management of horses in China remained monopolized by the bureaucratic group (Denis Sinor). The prohibition on the daily use of horses left these horse-breeding households deeply in debt, driving them to evade their conscription duties and even abandon their farmland. Only when fast donkey or horses are purchased at a fair price (sell to farmers after the war) or their owners serve as cavalrymen will the horses be properly cared for. official-operated enterprises not only makes the country frequently invaded by nomadic enemies, but also hinders the election of people with outstanding force strength. For example, the model of a large number of landlord soldiers being unable to kill a cavalryman led to the electoral system in ancient Mongolia (Morgan, David), where individual force was greater than landlord authority. Many technical force cavalry still existed in ancient Europe, and even gave rise to a specialized term: knights. The free trade and respect for the strong brought about by horse transportation opened up an arms race in Europe (Kelekna, Pita), and even in the 21st century, European police were widely using cavalry. On the contrary, the need for ignorant rule and bureaucratic enslavement prevented the popularization of horse plowing civilization in ancient China, leading to Empress Dowager Cixi, as a ruler, dying of illness and fleeing on the road. Even with cavalry killing landlords, taxation in the Qing Dynasty (not horse plowing) was still not ideal, but there will always be people who need to pay taxes. Poor people don't have the money to pay taxes, and businessmen don't want to pay taxes, but officer and the poor need money, so officer and the unemployer spontaneously attack business and advanced machines, which makes the development of business even slower.

**Mutual supervision and balanced economic equity incentive civilization**

The development of economic income follows a lognormal distribution, so the equity of enterprises can be divided into a lognormal distribution. Personal wealth is highly correlated with the stocks held, so in order to achieve full employment and reduce economic crises, the equity of enterprises should be granted to employees and stimulate their fighting spirit. This model, called equity incentives, is widely used by modern enterprises (Arbaugh, Larry). When distributing equity, it is also necessary to maintain a certain degree of rationality, such as allocating based on some models. Equity incentives represent the right to distribution, while money is a matter of shareholding ratio. Without equity incentives, net profits not converted into year-end dividends will not be taxed as equity dividends. If an enterprise does not redeem equity when an employee resigns, gains from the value-added of net asset will not be realized or subject to taxation. Debates and arguments are critical in any enterprise or nation, for you want others to believe that such practices are upright and promising. Lognormal distribution can better describe the "long tail feature" of income distribution, that is, the phenomenon where a small number of people occupy the high-income range and the income of the population on the left side of the median is lower than that of the main population (Colombia, Roberto). This also indicates that there will definitely be people who have to tighten their belts in society, but they should not be eliminated like a power-law distribution - too many people are eliminated, because humans are also valuable renewable resources. The mean and standard deviation in a log normal distribution can be understood as the wealth gap. We can adjust the wealth gap through the standard deviation $\sigma$ and $\mu$ to achieve sustainable growth in average wealth. Only by achieving a balance between mean, standard deviation and wealth gap can we achieve sustained development by stepping on the left foot and the right foot to ascend to the sky.

The development mode of human civilization profoundly affects its fate. Short sightedness and corruption led to the stagnation of civilization due to the prioritization of rule over production, and the oppression of the landlord class led by the emperor resulted in social impoverishment and knowledge gaps. In contrast, the horse plowing civilization promoted social progress through cavalry deterrence and free trade, while the European knight system fostered competition and innovation. The modern economy needs to draw on this wisdom and adopt a lognormal distribution equity incentive model to control the wealth gap while maintaining development momentum. Only by balancing efficiency and fairness, force and civilization, can sustainable social development be achieved. History has proven that stability that suppresses innovation will eventually decline, and moderate competition is the driving force behind civilization progress.

## Results

The horse plowing civilization not only improves production efficiency, but also suppresses unreasonable tax resistant groups, ensuring that no group is exempt from taxes. Modern economic research shows that there are profound mathematical laws between national income distribution and wealth structure. This article proves through rigorous mathematical deduction that national income follows a lognormal distribution, and its mathematical characteristics stem from the continuous compound interest growth mechanism; There is a significant positive correlation between personal wealth and shareholding ratio; Based on the lognormal distribution, the distribution of the number of people in a specific shareholding interval can be accurately calculated. These conclusions are not only supported by theoretical models, but also highly consistent with empirical data from various countries, providing a solid mathematical foundation for understanding wealth distribution and designing equity incentives. The following will be discussed in detail in two parts.

### The horse plowing civilization crushed the landlords

Proving the advantage of cavalry over landlord soldiers based on speed and killing efficiency. Cavalry speed $v_c$ ≈6m/s (charging speed of warhorses 20-25km/h); Infantry speed $v_i$ ≈1.5m/s (rapid army speed 5km/h); If the reaction time delay $\Delta t \approx 2s$ (human neural transmission and decision-making time), then the relative speed advantage of cavalry over infantry is $\Delta v = v_c$ $v_i = 4.5m/s$. When the charging distance d=100m: the arrival time of cavalry $t_c = d/v_c \approx 16.7s$, the effective defense time of infantry after reaction $t_i' = t_c - \Delta t \approx 14.7s$, and infantry can only move $d_i = v_i \times t_i \approx 22m$ during this time. Cavalry can break through 82% of infantry's defense depth ((100-22)/100) before establishing effective defense.

Use the Lanchester equation model to demonstrate the killing efficiency. dAdt =-βB, dB/dt =-αA. Where A is the number of cavalry (initial $A_0$=1); B is the number of landlord soldiers (initial B0=1000); α is the cavalry killing coefficient (including speed advantage); β is the infantry killing coefficient. Assuming: cavalry single attack efficiency $k_c$=0.1 (with a 10% probability of killing with a saber/spear); Infantry single attack efficiency $k_i$=0.0002 (hit rate of bow and arrow/spear against mobile cavalry); If the attack frequency ratio f=$v_c$/$v_i$=4, then the effective killing coefficient is: α=$k_c$ × f=0.4; β =$k_i$ =0.0002. At the end of the battle (A → 0), the remaining number of infantry is $B_f$=√($B_0^2$ -(α/β)$A_0^2$) ≈√($10^6$-2000) ≈999. Through the advantage of speed (Δv=4.5m/s) and killing efficiency (α/β=2000:1), it is proven that cavalry can theoretically concentrate their forces, attack with more and less, and maneuver to break through supply lines; Under actual battlefield conditions, one cavalry can effectively counter hundreds of landlord soldiers; The ancient battle example of "one hundred cavalry destroy a country" has mathematical rationality. The imperial-examination ranking will eventually turn into the force-examination ranking. This explains how cavalry units in history crushed landlord armed forces through mobility, dismantled their land control rights, and collected taxes. If horses cannot cultivate, a faster donkey can be used. Donkeys can not only defeat simple soldiers, but also be used for daily use. If transporting animals delivers an unwanted item to people who need it for exchange, both utility and use-value increase. At the same time, the civilization of horse plowing obeyed the rule of the strong and jointly voted to formulate laws, which did not result in the downfall of the people and the continuous evolution of the legal system.

**Using lognormal distribution to divide income**

(1) The mathematical proof of the logarithmic normal distribution of national income. If the random variable X>0 satisfies lnX~N (μ, $σ^2$) (i.e. logarithmically follows a normal distribution), then X follows a lognormal distribution, which is right skewed (long tail to the right, low income individuals to the left of the majority), suitable for describing the phenomenon of "positive variables and cumulative fluctuations", which is consistent with the empirical characteristics of national income distribution (a few high-income individuals, most low-middle income individuals). From the income growth mechanism to the lognormal distribution as a continuous compound interest growth model, assuming that income growth follows the "proportional growth" rule: the relationship between income $Y_t$ in a certain period and income $Y_{t-1}$ in the previous period is Yt=$Y_{t-1}$·(1+$g_t$), where gt is the independent and identically distributed growth rate (mean $μ_g$, variance $σ^2_g$). Take the logarithm of the equation: ln$Y_t$=ln$Y_{t-1}$+ln(1+$g_t$). If $g_t$ is small, approximately ln (1+$g_t$) ≈$g_t$, then ln$Y_t$ ≈ ln$Y_{t-1}$+$g_t$. After n iterations: ln$Y_t$≈ ln$Y_{t-1}$+$g_t$. According to the central limit theorem, when n is sufficiently large, the sum of independent and identically distributed variables approaches a normal distribution, i.e. lnYn~N (ln$Y_0$+nug, n$σ^2$), i.e. Yn follows a lognormal distribution. The high-income group approximately follows a Pareto distribution (P (Y>y) ∝$y^{-a}$), while the tail of the lognormal distribution asymptotically approaches $y^{[-(lny-μ)2:σ2lny]}$ at large values. When y is sufficiently large, it approximately decays in a power-law manner. After taking the logarithm of income data from various countries (such as the SCF survey in the United States and the household income survey in China), normality tests show that the logarithmized data is closer to a normal distribution, and the goodness of fit is significantly higher than that of directly fitting a normal distribution. From a mathematical derivation perspective, if income growth satisfies the assumptions of proportionality and independent and identically distributed distribution, the central limit theorem ensures that the logarithm of income tends towards a normal distribution, which is highly consistent with the characteristics of a right-handed income distribution and faster high-level economic growth in reality. Despite limitations such as thick tails and policy interventions, the lognormal distribution remains the fundamental model for characterizing national income distribution, and its mathematical simplicity and empirical effectiveness make it an important tool for economic analysis.

(2) Proof of a positive correlation between shareholding and wealth. Wealth W=αV+L (α is the shareholding ratio, V is the enterprise value, and L is the net value of other assets), taking the derivative of α yields dW:dα=V>0 (when V>0), proving a positive correlation between shareholding ratio and wealth. Considering dividends D=α·d·π and capital gains △V=α ($V_t$-$V_0$), it also indicates that equity returns increase wealth with increasing shareholding ratios. This indicates that in reality (Employees contribute the initial investment amount or net asset investment amount), equity is highly correlated with the initial investment amount, equity appreciation, and equity dividends. Employees cannot be taxed on three types of income unless they actually receive all three.

It can be seen from the above that all employees' income is divided into three parts after purchasing stocks: salary, equity dividends and equity appreciation. If salaries are set at an extremely low level close to zero, product prices will only cover raw material costs. Since zero salary is impractical, the question arises as to what an appropriate salary level should be. A sound compensation structure needs to secure basic livelihoods, offer individual incentives as well as collective incentives: salaries cover daily living expenses, bonuses reward individual performance, dividends drive corporate performance incentives, and equity appreciation compensates for service tenure, namely the net asset difference between what employees paid to buy company shares when they joined, and what the company pays to buy those shares back when they leave.. Equity appreciation is generally realized only after decades, resulting in difficulties in supervision; nevertheless, salaries, bonuses and dividends (dividends=net profit, after deducting 10% or 20% surplus reserve, becomes the dividends available for distribution to investors) are all distributed within one year, so their proportions can be adjusted to roughly 1:1:1. The ratio of dividends and bonuses fluctuates around wages. None of the three components is permitted to hold a dominant share, in order to prevent problems including inability to sustain livelihood without basic salary, idling on dividends alone, and excessive focus on short-term bonuses.

(1)Calculate the number of individuals in the shareholding range under a lognormal distribution. Calculate the highest value as μ (probability density is the proportion, x-axis is the number of shares held). Given that the number of shares held by employees S follows a lognormal distribution, i.e. lnS~N (μ, $\sigma^2$), where μ=lnS; σ reflects the degree of distribution dispersion (related to the definition of "standard deviation=wealth gap" in the paper).

The above problem requires calculating the quantile difference of the log-normal distribution. The quantile difference of the cumulative probability density is essentially the quantile difference of the ratio P (0<x<b). The all number of virtual employees is M; the all number of real employees is N, and M > N. The number of virtual employees at the front end is s.

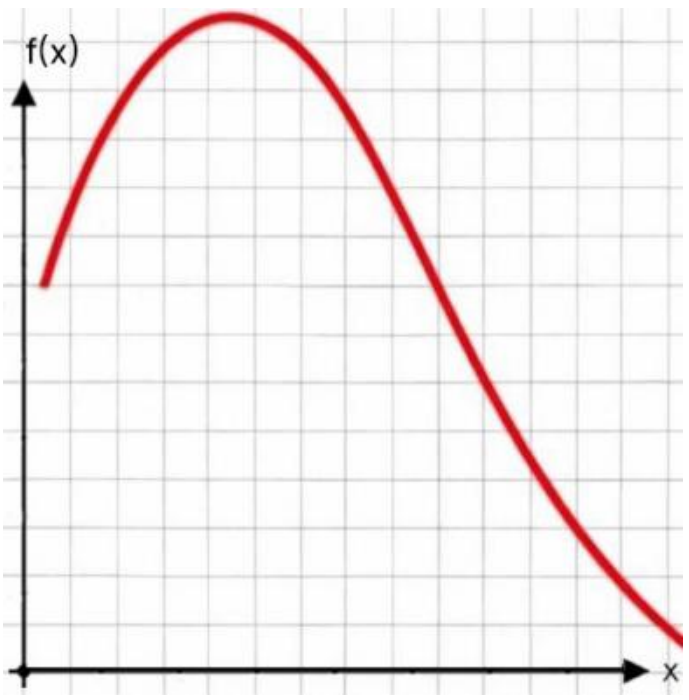

Figure 1. The horizontal axis can be interpreted as the share-holding number; a larger value of x corresponds to greater personal wealth. Rotating the graph will be more intuitive, as you can see the x-axis changing to the y-axis.

The cumulative probability function of the log-normal distribution over the interval [0, x] takes a value equal to 1÷M, and all values of x lie on quantile points. Starting from a approaching 0, the cumulative probability increment (i.e., probability area or proportion) of the i-th interval is $F_{xi}$. As illustrated in the figure, three virtual employees exist before $x_1$, and one virtual employee appears after $x_N$. The points corresponding to real employees are successively $x_1$, $x_2 \ldots x_N$. The proportion corresponding to $x_i$ equals (s+i)÷M. Accordingly, the equation for cumulative probability density corresponding to the i-th scale in practical scenarios is:

$$P\left(0<X<x_i\right) = \frac{s+i}{M} = F\left(x_i\right) = \Phi\left(\frac{\ln x_i - \mu}{\sigma}\right)$$

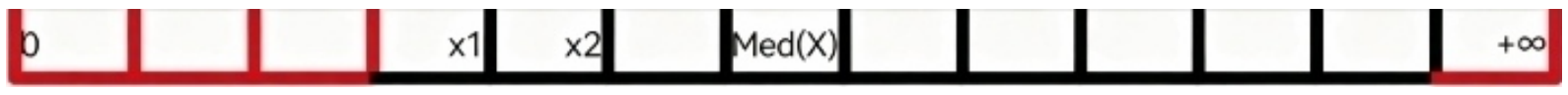

Figure 2. This diagram approximates the curve with straight lines to illustrate that the proportions among excluded employees follow a proportional relationship. Black line segments denote the scales of real employees. There are 13 virtual employees and 9 real employees, with 3 virtual employees on the left.

The number of employees excluded on the left and right ends is equal, with s=(M-N)÷2 employees excluded at each end. The two tails of the distribution expand or contract simultaneously. This process can be regarded as extending outward on both sides centered on the median line Med(X). This equalization model accounts for the following scenario: management threatens to dismiss voters and proposers to coerce them into voting in favor of raising managers' salaries, while ordinary employees' wages remain unchanged.

$$\frac{\frac{M-N}{2}+i}{M} = \Phi\left(\frac{\ln x_i - \mu}{\sigma}\right); \qquad \frac{\ln x_i - \mu}{\sigma} = \Phi^{-1}\left(\frac{\frac{M-N}{2}+i}{M}\right);$$

$$\ln x_i = \mu + \sigma\Phi^{-1}\left(\frac{\frac{M-N}{2}+i}{M}\right); \qquad x_i = \exp\left(\mu+\sigma\Phi^{-1}\left(\frac{\frac{M-N}{2}+i}{M}\right)\right)$$

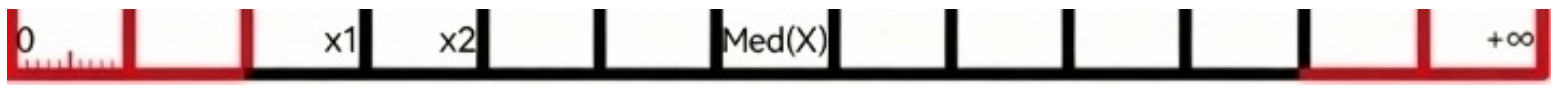

Figure 3. This formula features favorable fairness and deserves wide application. Suppose there are a total of 13 virtual employees, with two eliminated at each end, leaving 9 real employees. The value of the cumulative distribution function corresponding to each employee equals 1/13.

Only when this cumulative probability is exactly equal to 1÷N does it correspond to one employee; but the quantile difference is still the width of the wage interval in which this employee is located. The quantile difference of the i-th employee. Let s=(M-N)÷2, substitute $x_i$ into the quantile function of the log-normal distribution, and we obtain the quantile difference of the i-th employee. The formula is stated as follows:

$$\Delta_i = x_i - x_{i-1} = \exp\left(\mu+\sigma\Phi^{-1}\left(\frac{M-N+2i}{2M}\right)\right) - \exp\left(\mu+\sigma\Phi^{-1}\left(\frac{M-N+2(i-1)}{2M}\right)\right)$$

Summation formula for quantiles of N employees. It can be added after retaining two decimal places

$$\Sigma_{i=1}^{N}\left(x_i - x_{i-1}\right) = x_N - x_0 = \exp\left(\mu+\sigma\Phi^{-1}\left(\frac{M+N}{2M}\right)\right) - \exp\left(\mu-\sigma\Phi^{-1}\left(\frac{M-N}{2M}\right)\right)$$

Each quantile proportion, namely the proportion of all wages occupied by $x_i$, is cancelled out $e^u$

$$r_i = \frac{\exp\left(\sigma\Phi^{-1}\left(\frac{M-N+2i}{2M}\right)\right) - \exp\left(\sigma\Phi^{-1}\left(\frac{M-N+2(i-1)}{2M}\right)\right)}{\exp\left(\sigma\Phi^{-1}\left(\frac{M-N}{2M}\right)\right) - \exp\left(\sigma\Phi^{-1}\left(\frac{M-N}{2M}\right)\right)}$$

$\Phi^{-1}$ (p)=$z_p$. Therefore, the final formula for the equity proportion of the i-th-ranked employee is:

$$r_i = \frac{\exp\left(\sigma z_{\left(\frac{M-N+2i}{2M}\right)}\right) - \exp\left(\sigma z_{\left(\frac{M-N+2(i-1)}{2M}\right)}\right)}{\exp\left(\sigma z_{\left(\frac{M-N}{2M}\right)}\right) - \exp\left(\sigma z_{\left(\frac{M-N}{2M}\right)}\right)}$$

This characteristic reveals the rigor of this distribution system—rankings matter greatly. This will trigger internal competition among employees at the same tier. Hence, salary equalization is required for certain tiers. Equal pay for equal positions can also be observed in reality; for instance, wages for some basic operators are uniform. First, calculate the theoretical salaries of employees No.3 to No.7. Compute the average value of the sum of these five employees' salaries and adopt it as their unified wage.

$$w_{avg} = (w_{xi} + w_{x(i+1)} + w_{x(i+2)} + \ldots + w_{xq}) \div (q-i+1)$$

**Parameter description**:

σ: The standard deviation of the logarithmic part of the log-normal distribution (the core parameter determining distribution skewness), adjustable.

M: All virtual employees. M denotes the number of virtual employees with an adjustable ratio, adjustable. It determines the overall base of salary.

s: s and (M-N)÷2 are essentially proportional scales rather than discrete numerical scales. That is, s may represent one employee, or fractional equivalents such as 0.5 employees, 0.3 employees, and so forth.

μ: The mean value of the logarithmic part of the log-normal distribution, unadjustable and nearly useless.

N: Total real employees. The quantile corresponding to each interval is M-(M-N÷2)×2=1÷N. The number of selected quantiles. Summation runs from i=1 to i=N. N stands for the number of real-world employees and is fixed.

$w_{x1}$: Subsistence-level wage, set at 3000, adjustable (reflecting the harsh nature of this model, which requires equalization).

$r_{x1}$: The proportion of the subsistence-level wage. It varies constantly, determined by the number of total virtual employees and front-end virtual employees; standard deviation. Adjust these three variables to alter $r_{x1}$ and keep the subsistence-level wage around 3000. Given the all wage w and $w_{x1}$, we have $r_{x1}=w_{x1}\div w$.

$w_{xn}$: The proportion of the highest-paid employee's wage, determined by whether managers resign. It is governed by the number of total virtual employees and front-end virtual employees; standard deviation, and ultimately decided by voters.

q: The ranking of employees that need to be averaged. Equalizing the average salaries of employees at the same level aims to reduce internal conflicts. This prevents colleagues from engaging in infighting over performance rankings and pay allocation ratios while working.

Under the contribution-based ranking system, employee ranks rise gradually from left to right. If managers are incompetent, voting reverses the ranking order, with employee ranks declining progressively as shown in Figure 2. This means the managers must resign immediately. Most people will not reverse the ranking if managers achieve satisfactory performance.

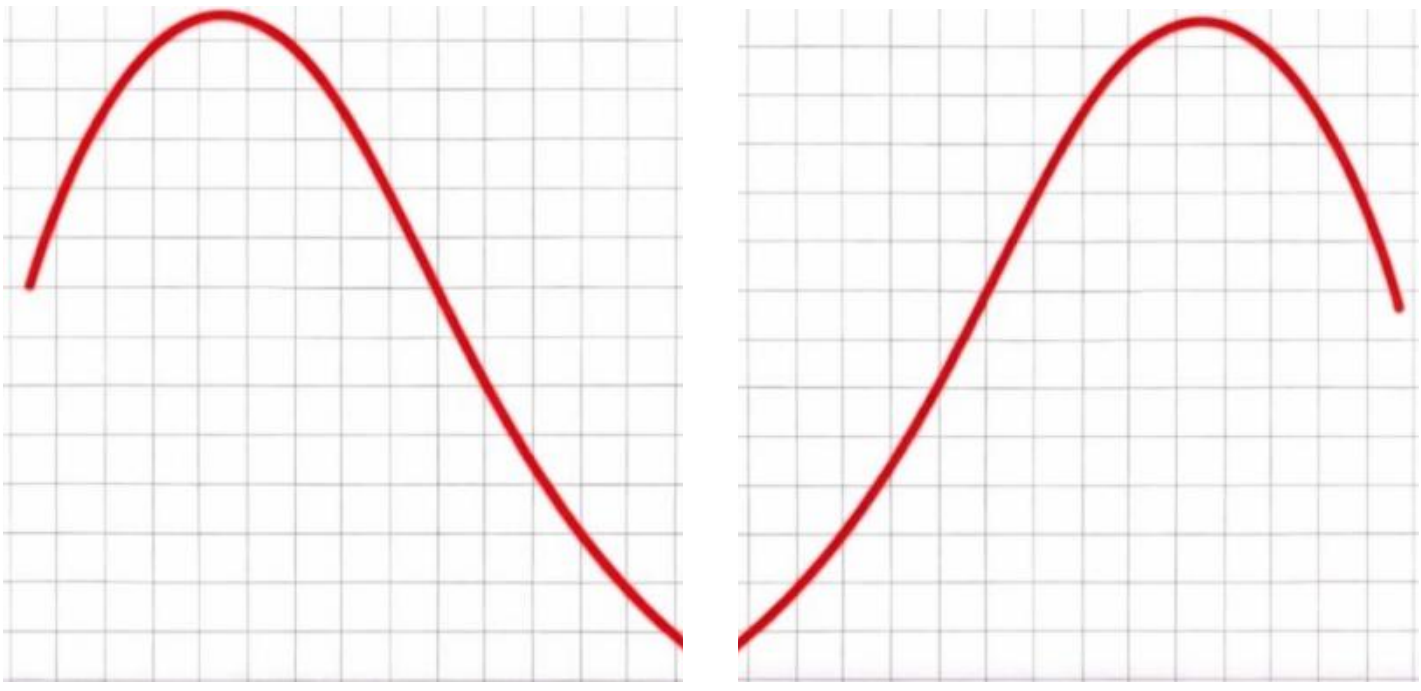

Figure 4. Assuming a total of 16 employees, with the two at each end removed, there are 12 actual employees remaining. The CDF for each employee is 1/16. It means that the last place will become the first place.

In this model, the quantile $b_i = e^{\mu} \times c_i$, so $e^{\mu}$ cancels out in the ratio calculation, meaning the relative allocation proportion has nothing to do with the mean μ and is determined only by σ , M, N and i; μ merely scales the overall absolute wage level uniformly for all employees without altering the internal pay structure, ranking or wealth gap, which is why it appears irrelevant in core proportional and structural analysis, serving only as a base adjustment factor.

Here arises a question. Since specific-value selection cannot be determined by voting, how should we make the choice? The optimal outcome is to adjust the magnitude of each data point to turn them into individual voting proposals. For instance, σ = 10 and M = 30 constitutes one allocation scheme (hierarchy is a secondary concern), while σ = 30 and M = 120 forms an equity-allocation scheme for calculating concrete wages. σ can be multiples of 10 such as 10, 20, 60 or multiples of 30 such as 60, 90, 120. To reduce the complexity of data arrangement, we need to minimize the number of parameters that determine the allocation model. More parameters mean greater-adjustable degrees of freedom for the model and more complicated proposals. For example, two parameters can be selected: standard deviation, virtual total number of employees. Choosing two parameters is analogous to the three-body problem in physics. Since the three-body problem lacks a general analytical solution, it implies that this model can generate infinite variations, while keeping the number of parameters manageable. Multiples are first determined by voting with values kept as small as possible so that the enterprise can generate numerous proposals. Besides, every ordinary employee excluding the highest-ranked staff may draft a voting slip containing n calculation results and post it on the noticeboard for public disclosure. The meaning is that one piece of paper can write n plans. Employees cast anonymous votes for these posted paper proposals, and the proposal with the most votes wins. Winners of the proposal election shall receive considerable economic gains, while elected top-level managers shall hold senior-management positions. In this way, the group that selects the distribution scheme is separated from the group that elects the administrators, realizing mutual checks and balances between the two mechanisms.

As for who shall hold the management rights, managers should be elected through a vote based on the principle of one person, one vote. Why do we do this? Because managers hold the power of position allocation, and employees must have the right to decide who the managers are. Management does not necessarily have to exist--so long as long-term equity incentives are sufficient and distribution is reasonable, employees will work spontaneously without the need for intensive management. This ensures that the enterprise does not operate under a chieftain-allocated system, since employees are equally entitled to decide who their managers are. The employees elect the managers, model structure (expected value, standard deviation, the difference or virtual multiple between the minimum wage and the maximum wage), and the managers determine the employees’ position; the two sides check and balance each other.If it is not suitable, employees may not vote to elect the allocation plan and the manager who formulated the plan (shareholder power). A utopia where everyone enjoys equal equity returns does not exist, because while employees are granted voting rights, it is imperative to guarantee the returns of shareholders. The term of office lasts one year, which is sufficient to reflect the growth in operating revenue. A higher growth rate also qualifies one for re-election. People are born neither good nor evil, but sometimes good and sometimes evil. Because spear-shield=0, so even if the power of managers and employees grows or declines, it can be beneficial or harmful to the organization, represented by ±. So $\pm a(1\pm x)^t \pm b(1\pm y)^t = \pm c(1\pm z)^t$. If the result c is positive value, the organization exists. If both are equal and positive, it possesses the greatest organizational strength c. This is precisely for the purpose of mutual restraint and rising.

**Explanation**

Why use a structure like the lognormal distribution, which is "large in the middle and small at both ends"? Because in an enterprise, for example, temporary workers are also employees serving the enterprise, so these employees cannot all be outsourced. If outsourcing were allowed, then could the lower-level employees engaged in the company's main business also be outsourced? Who has the decision-making power over which outsourcing company to use? Undoubtedly, it is the managers, so it is easy for managers to take kickbacks. If managers obtain money through embezzlement, does that conform to the principle of efficiency?

Why must China use layer-by-layer outsourcing? After foreign investors' factories and machinery, which cost enormous amounts, have been put in place, foreign investors, in order to avoid being deliberately obstructed by local officials—in small places, if this official does not take a cut, another official will—use outsourcing companies and subcontracting companies in which bureaucrats hold hidden shares, causing enterprises to subcontract labor layer by layer, with employees producing low-quality products at low wages. Shareholders and bureaucrats collude to exploit employees. If layer-by-layer subcontracting is carried out, for example, if a project is outsourced four or five times, then the wages of the workers at the bottom are only 2% of the project payment. Can such bottom-level workers provide good service?

Why use such a simple equation? Because the simpler and more crude it is, the less likely the truth is to fail in the process of operation. If there are too many parameters, it is easy for management to manipulate them. Using an equation, even without a computer, one can calculate it—keeping two decimal places before accumulation—so that voting distribution plans can be carried out even in the most backward areas. If there are too many parameters and the wage ratio cannot be calculated clearly on one sheet of paper, then everyone can formulate voting plans, and it becomes difficult to implement them. Therefore, the parameters have to be simplified, but only two. Later, it can also be suitable to add one or two parameters.

Why can't distribution be equal? Because different employees obtain different equity net asset appreciation. Equal distribution is also unfair. Suppose ordinary employees collectively obtain ownership of the company; out of their own interests, they will choose to lay off workers, because new members will divide the existing members' benefits, and only when the contribution of newcomers exceeds the current per capita level is it worth absorbing them. Use real examples as evidence: in developed countries, some professional groups' unions, such as doctors' unions, lobby to restrict the number of new entrants to the profession.

Why use a contribution ranking system? The contribution ranking system is a kind of performance contribution ranking synthesized from various contributions. Some enterprises lean toward sales, and some lean toward production, so it cannot be determined what kind of contribution should be used; therefore, performance ranking is used. The survival of the fittest within an enterprise is also unavoidable, so a performance ranking system must be used. An enterprise must maintain progress internally to counter external technological progress.

Why elect one manager to manage everyone's performance ranking? For the sake of the enterprise's speed of risk response and to avoid delays caused by wrangling. The manager serves for one year and is professional. After stepping down, he or she still has to continue working in the enterprise and will not drain the pond to catch all the fish.

The minimum and maximum constraints are determined by the proportions removed from the front end and the back end of the lognormal distribution. I did consider 3 times and 5 times, but when should it be 3 times and when should it be 5 times? Some people say it should be 4 times, and some say 6 times. What should be done then? Whose opinion should be followed? Therefore, I chose to follow the quantile growth pattern of the lognormal distribution. The main fear with too many parameters is that management will exploit the confusion for profit.

## Discussions

The essence of civilization upgrading is the coordinated evolution of production efficiency, power structure, and benefit distribution. The key to achieving balance from the closure of human cultivated civilization to the breakthrough of horse cultivated civilization, and then to modern equity incentives lies in institutional design. The mutual promotion of force technology and economic models: the cavalry speed advantage ($\Delta v=4.5m/s$) and the Lanchester equation prove that military displacement needs to be combined with economic mechanisms (such as free trade promoted by European knights) to form a positive cycle, while ancient Chinese official horse politics became a shackle of rule due to institutional rigidity. The mathematical regulation of wealth distribution: Lognormal distribution adjusts through mean and standard deviation, which can stimulate innovation (positive correlation between equity and wealth $dW/d\alpha=V>0$) and suppress extreme differentiation. Its tail characteristics and Venetian mixed rule mechanism both reflect the importance of "institutional elasticity". Dynamic management of conflict thresholds: Cases such as the Glorious Revolution in Britain have shown that moderate internal pressure can drive reform, avoid cyclical economic crises,while extremism (such as wealth concentration under power-law distribution) will inevitably lead to collapse. Contemporary governance requires the construction of a "historical parameter mathematical model" framework, using tools such as equity incentives to achieve "controllable quenching" and avoid falling into the civilization trap of "advanced technology but lagging governance".

## Method

Historical analysis method: By sorting out the evolution of different civilizations such as ancient China, Europe, and the Mongol Empire, typical characteristics of human cultivated civilization and horse cultivated civilization are extracted. By comparing historical events such as the collapse of the Roman Empire and the Glorious Revolution in England, we can summarize the balance law between economic benefit distribution and power suppression. Well-allocated small teams can evolve into enterprises, providing historical references for modern governance.

Mathematical modeling method: Construct a Lanchester equation model, introduce parameters such as cavalry speed (6m/s) and infantry speed (1.5m/s), quantitatively analyze the military advantage of cavalry over landlord soldiers, and verify the mathematical rationality of "one thousand cavalry breaking ten thousand". Based on the theory of lognormal distribution, a mathematical model of national income distribution is derived, and wealth distribution is adjusted through number virtual employees (M), number of real employees (N) and standard deviation ($\sigma$). Combined with the scenario of corporate equity incentives, the distribution of the number of people in the shareholding range is calculated, providing a quantitative tool for narrowing the wealth gap. Establish a correlation model between shareholding and wealth ($W=\alpha V+L$), prove the positive correlation between equity ratio and wealth through differentiation, and verify the economic logic of equity incentives by combining dividend and capital gains mechanisms.

Interdisciplinary analysis method: Integrating economic, force, and historical theories, transforming political practices such as the electoral system of the Venetian Republic and the Chinese imperial examination system into institutional design parameters, embedding them into mathematical models to optimize the selection mechanism for rulers, and achieving cross validation of historical experience and mathematical analysis. Using the cause and effect of logic, the spear and shield of dialectics to integrate various disciplines, thereby promoting social progress in the struggle. Say's Law holds that supply creates demand, while Keynes' Law holds that demand creates supply. But if demand and supply want to continue to grow and be equal, they must create each other and spiral upwards.

## Future predictions and expectations

The future of civilization will belong to the group that can transform "historical databases" into "evolutionary algorithms" - whether it is countries or enterprises. Only by establishing an adaptive system of "mathematical model prediction+institutional flexibility adjustment" can we avoid falling into the trap of "advanced tools but stagnant civilization" in the era of technological explosion, and achieve a qualitative change from "passive adaptation" to "active evolution". The legitimacy of civilization needs to be found in history and the hearts of the people, otherwise the people will turn from sheep to wolves. Senior officials shall be duty-bound to formulate a simple and understand-to-efficient distribution mechanism, instead of allowing the exploitation chain where senior officials prey on junior ones, junior officials prey on the people, and the people overthrow senior officials—for such a vicious cycle brings no progress at all.

## The contradiction between wealth ratio gap and cash issuance

The founding mechanism of the Bank of England: In 1694, the King of England took out a private loan of 1.2 million pounds in silver coins, pledging tax revenue to cover interest payments. The annual cost to private creditors included 8% interest plus a 4,000-pound management fee, payable primarily in silver; private lenders would not accept worthless paper promises. The creditors received negotiable credit instruments called banknotes, which later evolved into paper money widely used for commodity transactions. The state secured silver coins to maintain fiscal operations, resulting in a mutually beneficial arrangement. Wiping out the huge sovereign debts of modern nations would eliminate paper currency and trigger deflation.

This system has a fundamental logical defect: Because bond interest is financed by government tax revenue, the total amount of debt issuance is limited by taxable income. The limit is determined by ordinary people’s tolerance for monetary expansion. Since money merely represents proportional wealth, paper currency can theoretically be printed without bound, and this excess issuance ultimately shows up in commodity prices. When wealth inequality becomes severe and ordinary citizens own only a minimal share of total wealth, two scenarios unfold. Ordinary people will either lack money for consumption and tax payments, or people opt to disengage from work and forego having children amid harsh living conditions.. In both cases, the public can no longer support society’s debt burden. Tax revenue collapses, new currency cannot be issued. The excessively low relative prices of grain and tools will also push dignified laborers to adopt barter, which will ultimately bring the entire social system to collapse, and the whole social system breaks down. Consequently, humanity must maintain a cyclic equilibrium between wealth disparity and currency creation.